\documentstyle[12pt,psfig]{article}
\begin{document}
\bibliographystyle{unsrt}
\noindent
\begin{center}
{\bf The Structure and Dynamics of Sodium Disilicate}

J. Horbach and W. Kob\footnote{Author to whom correspondence should 
be addressed to. E--mail: kob@moses.physik.uni-mainz.de,
http://www.cond-mat.physik.uni-mainz.de/\~{ }kob} \\
Institute of Physics, Johannes Gutenberg University, Staudinger Weg 7,\\
D-55099 Mainz, Germany
\end{center}

\vspace*{7mm}
\par
\noindent
\begin{center}
\begin{minipage}[h]{122mm}
We investigate the structure and dynamics of sodium disilicate by means
of molecular dynamics computer simulation. We show that the structure
is described by a partially destroyed tetrahedral SiO$_4$ network and a
spherical super structure formed by the silicon and sodium atoms. The
static structure factor of our simulation is in very good agreement
with one from a neutron scattering experiment. For $1008$ particles we
find strong finite size effects in the dynamics which are due to the
missing of modes contributing to the boson peak.
\end{minipage}
\end{center}

\vspace*{5mm}
\par
\noindent

\section{Introduction}

In recent years several molecular dynamics computer simulations have
been done in order to investigate the structure and dynamics of sodium
silicate melts and glasses (Smith, Greaves, and Gillan 1995, Cormack
and Cao 1997). By using the potential proposed by Vessal, Amini,
Fincham and Catlow (1989) these authors have found that the structure
of systems like, e.g., sodium disilicate (SDS) is characterized by a
microsegregation in which the sodium atoms form clusters of a few atoms
between bridged SiO$_4$ units. In order to see whether this somewhat
surprising result is reproduced also by a different model from the
one of Vessal {\it et al.}~(1989) we have performed simulations of SDS
using a different potential (discussed in detail below). In addition to
the investigation of the structure we also study the dynamical
properties of SDS in order to see whether the finite size effects that
have been observed in pure silica (Horbach, Kob, Binder, and Angell
1996, Horbach, Kob, and Binder 1999a, and Horbach, Kob, and Binder
1999b) are present in SDS as well.

\section{Model}

The model potential we use to describe the interactions between the
ions in SDS is the one proposed by Kramer, de Man, and van Santen
(1991) which is a generalization of the so called BKS potential (van
Beest, Kramer, and van Santen 1990) for pure silica. It has the
following functional form:
\begin{equation}
\phi(r)=
\frac{q_{\alpha} q_{\beta} e^2}{r} + 
A_{\alpha \beta} {\rm e}^{-B_{\alpha \beta}r} -
\frac{C_{\alpha \beta}}{r^6}\quad \alpha, \beta \in
[{\rm Si}, {\rm Na}, {\rm O}].
\label{eq1}
\end{equation}
Here $r$ is the distance between an ion of type $\alpha$ and an ion of
type $\beta$. The values of the parameters $A_{\alpha \beta}, B_{\alpha
\beta}$ and $C_{\alpha \beta}$ can be found in the original
publication. The potential (\ref{eq1}) has been optimized by Kramer
{\it et al.}~for zeolites, i.e.~for systems that have Al ions in 
addition to Si, Na and O. In that paper the authors used for silicon
and oxygen the {\it partial} charges $q_{{\rm Si}}=2.4$ and 
$q_{{\rm O}}=-1.2$, respectively, whereas sodium was assigned its real
ion charge $q_{{\rm Na}}=1.0$. With this choice charge neutrality is
not fulfilled in systems like SDS. To overcome this problem we
introduced for the sodium ions a position dependent charge $q(r)$
instead of $q_{{\rm Na}}$,
\begin{equation}
  q(r)= \left\{
  \begin{array}{l@{\quad \quad}l}
   0.6 \left( 1+
   \ln \left[ C \left(r_{{\rm c}}-r \right)^2+1 \right] \right) & 
   r < r_{{\rm c}} \\
   0.6  & r \geq r_{{\rm c}}
   \end{array} \right. \label{eq2}
\end{equation}
which means that for $r \ge r_{{\rm c}}$ charge neutrality is valid
($q(r)=0.6$ for $r \ge r_{{\rm c}}$). Note that $q(r)$ is continuous at
$r_{{\rm c}}$. We have fixed the parameters $r_{{\rm c}}$ and $C$ such
that the experimental mass density of SDS and the static structure
factor from a neutron scattering experiment (see below) are reproduced
well. From this fitting we have obtained the values $r_{{\rm
c}}=4.9$\AA~and $C=0.0926$\AA$^{-2}$. With this choice the charge
$q(r)$ crosses smoothly over from $q(r)=1.0$ at $1.7$ \AA~to $q(r)=0.6$
for $r \ge r_{{\rm c}}$.

The simulations have been done at constant volume with the density of
the system fixed to $2.37 \, {\rm g}/{\rm cm}^3$. The equations of
motion were integrated with the velocity form of the Verlet algorithm
and the Coulombic contributions to the potential and the forces were
calculated via Ewald summation. The time step of the integration was
$1.6$ fs. In this paper we investigate the properties of SDS in the
liquid state at $T=2100$ K and in the glass state at $T=300$ K. The
equilibration time at $T=2100$ K was two million time steps thus
corresponding to a real time of $3.5$ ns. At this temperature we
simulated systems with $N=1008$ and $N=8064$ particles. In order to
improve the statistics two independent runs were done for the large
system and eight independent runs for the small system. The glass state
was produced by cooling the system from equilibrated configurations at
$T=1900$~K with a cooling rate of $1.16 \cdot 10^{12} \, {\rm K}/{\rm
s}$. The pressure is $4.5$ GPa at $T=2100$ K and $0.96$ GPa at $T=300$
K.

\section{Results}

\begin{figure}[f]
\psfig{file=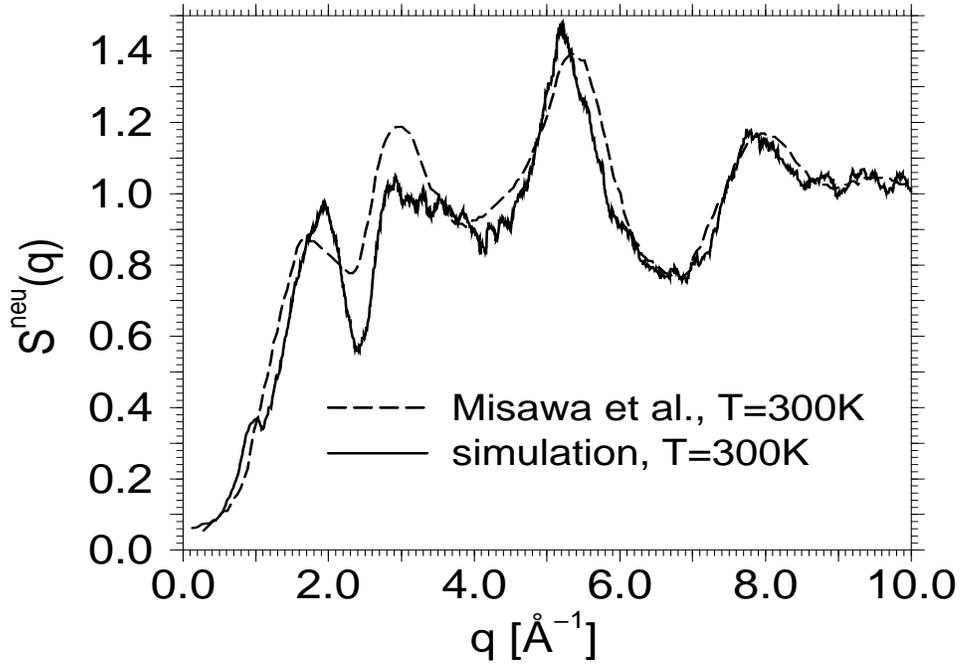,width=13cm,height=9.5cm}
\caption{Comparison of the static structure factor $S^{{\rm neu}}(q)$
	 from our simulation (solid line) with the experimental
	 data of Misawa {\it et al.}~(1980) (dashed line). \label{fig1}}
\end{figure}

In order to demonstrate that our model is able to reproduce the
structure of real SDS very well we compare the static structure factor
$S^{{\rm neu}}(q)$ with the one from a neutron scattering experiment by
Misawa, Price, and Suzuki (1980). To calculate $S^{{\rm neu}}(q)$ one
has to weight the partial structure factors from the simulation with the
experimental coherent neutron scattering lengths $b_{\alpha}$ ($\alpha
\in [{\rm Si, Na, O}]$):
\begin{equation}
   S^{{\rm neu}}(q) = 
   \frac{1}{\sum_{\alpha} N_{\alpha} b_{\alpha}^2}
   \sum_{kl} b_k b_l 
   \left< {\rm e}^{i {\bf q} \cdot [ {\bf r}_k - {\bf r}_l ]} 
   \right> .  \label{eq3}
\end{equation}
The values for $b_{\alpha}$ are $0.4149 \cdot 10^{-12}$ cm, $0.363 \cdot
10^{-12}$ cm and $0.5803 \cdot 10^{-12}$ cm for silicon, sodium and
oxygen, respectively. They are taken from Susman, Volin, Montague, and
Price (1991) for silicon and oxygen and from Bacon (1972) for sodium.
Fig.~\ref{fig1} shows $S^{{\rm neu}}(q)$ from the simulation and the
experiment at $T=300$~K. We see that the overall agreement between
simulation and experiment is good. For $q > 2.3$ \AA$^{-1}$, which
corresponds to length scales of next nearest Si--O and Na--O neighbors,
the largest discrepancy is at the peak located at $q=2.8$~\AA$^{-1}$
where the simulation underestimates the experiment by approximately
$15$\% in amplitude. Very well reproduced is the peak at
$q=1.7$ \AA$^{-1}$, which is called the first sharp diffraction peak,
and which is a prominent feature in pure silica as well. In silica this
peak arises from the tetrahedral network structure since the
length scale which corresponds to it, i.e.~$2\pi/1.7 \; {\rm \AA}^{-1}=
3.7$ \AA, is approximately the spatial extent of two connected SiO$_4$
tetrahedra. From the figure we recognize that this structure is 
partly present in SDS also. The peak at $q=0.95$ \AA$^{-1}$ is not 
present in the experimental data which might be due to the fact that in
this $q$ region the experimental resolution is not sufficient. By
looking at the coordination number distributions, discussed below, we
see that the peak at $q=0.95$ \AA$^{-1}$ is related to
a super structure which is formed by the sodium and silicon atoms. In
agreement with this interpretation the length scale corresponding to
this peak, i.e.~$2 \pi/0.95 \; {\rm \AA}^{-1}=6.6$ \AA, is two times
the mean distance of nearest Na--Na or Na--Si neighbors.

\begin{figure}[f]
\psfig{file=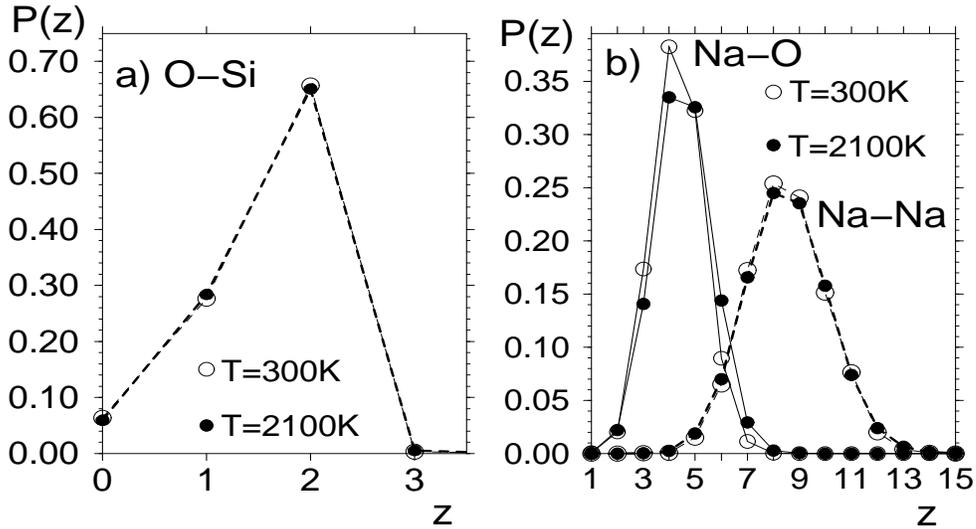,width=13cm,height=9.5cm}
\caption{Distributions of the coordination number for SDS at $T=300$ K
	 (open circles) and $T=2100$ K (filled circles), 
	 a) $P_{{\rm O-Si}}(z)$, b) $P_{{\rm Na-O}}(z)$ and
	 $P_{{\rm Na-Na}}(z)$. \label{fig2}}
\end{figure}

The coordination number distribution $P_{\alpha \beta}(z)$ for
different pairs $\alpha \beta$ gives the probability that an ion of
type $\alpha$ has exactly $z$ nearest neighbors of type $\beta$.  By
definition two neighboring atoms have a distance from each other which
is less than the location of the first minimum $r_{{\rm min}}$ of the
corresponding partial pair correlation function $g_{\alpha \beta}(r)$.
From the functions $g_{\alpha \beta}(r)$ we find for $r_{{\rm min}}$
the values $3.6$~\AA, $5.0$~\AA, $2.35$~\AA, $5.0$~\AA, $3.1$~\AA, and
$3.15$~\AA~for the Si--Si, Si--Na, Si--O, Na--Na, Na--O, and~O--O
correlations.  We note that $P_{{\rm Si-O}}(z)$ is larger than $0.99$
for $z=4$ at $T=2100$ K and $T=300$ K which means that nearly every
silicon atom is four fold coordinated with oxygen atoms forming a
SiO$_4$ tetrahedron. Some of the distribution functions are shown in 
Fig.~\ref{fig2}. We recognize from Fig.~\ref{fig2}a that about
$65$\% of the oxygen atoms are bridging oxygens between two tetrahedra
($P_{{\rm O-Si}}(z=2) \approx 0.65$), and that about $28$\% of the
oxygen atoms form dangling bonds ($z=1$) with corresponding silicon
atoms. In the neighborhood of these dangling bonds sodium atoms are
located. This means that the sodium atoms partly destroy the
(disordered) tetrahedral network which is the structure for pure
silica. A significant number of oxygen atoms have no silicon atoms, but
only sodium atoms as direct neighbors. In Fig.~\ref{fig2}b we show
$P_{{\rm Na-Na}}(z)$ at $T=2100$ K and $T=300$ K, and we see that
essentially the two distributions coincide with a
a mean value between $z=8$ and $z=9$.  Basically the
same distribution is found for $P_{{\rm Na-Si}}(z)$.  Therefore, every
sodium atom is surrounded by $8$--$9$ other sodium atoms and $8$--$9$
silicon atoms. Since the mean distance between Na--Na and Na--Si
neighbors is approximately the same, i.e.~about $3.3$ \AA, we can
conclude that sodium and silicon atoms form a spherical super structure
in which every sodium atom is surrounded by a first shell of 
oxygen atoms and a second shell of silicon and sodium atoms. In
Fig.~\ref{fig2}b we have also included $P_{{\rm Na-O}}(z)$ and we
recognize that every sodium atom has on average about $4$--$5$ nearest
oxygen neighbors.  There is again no essential difference between
$P_{{\rm Na-O}}$ for $T=2100$ K and $P_{{\rm Na-O}}$ for $T=300$ K,
although the pressure is about a factor $4.5$ higher at $T=2100$ K.
This means that the structural features which we observe at $T=2100$~K
are not formed due to the relatively high pressure.  Nevertheless, we
emphasize that the small difference between $P(z)$ in the liquid and in
the glass state is partly due to the high cooling rate of about
$10^{12} \; {\rm K}/{\rm s}$ we used to produce the structures at
$T=300$ K.  A careful analysis of the cooling rate effects for SDS will
be presented elsewhere (Horbach, Kob, and Binder 1999c).

\begin{figure}[f]
\psfig{file=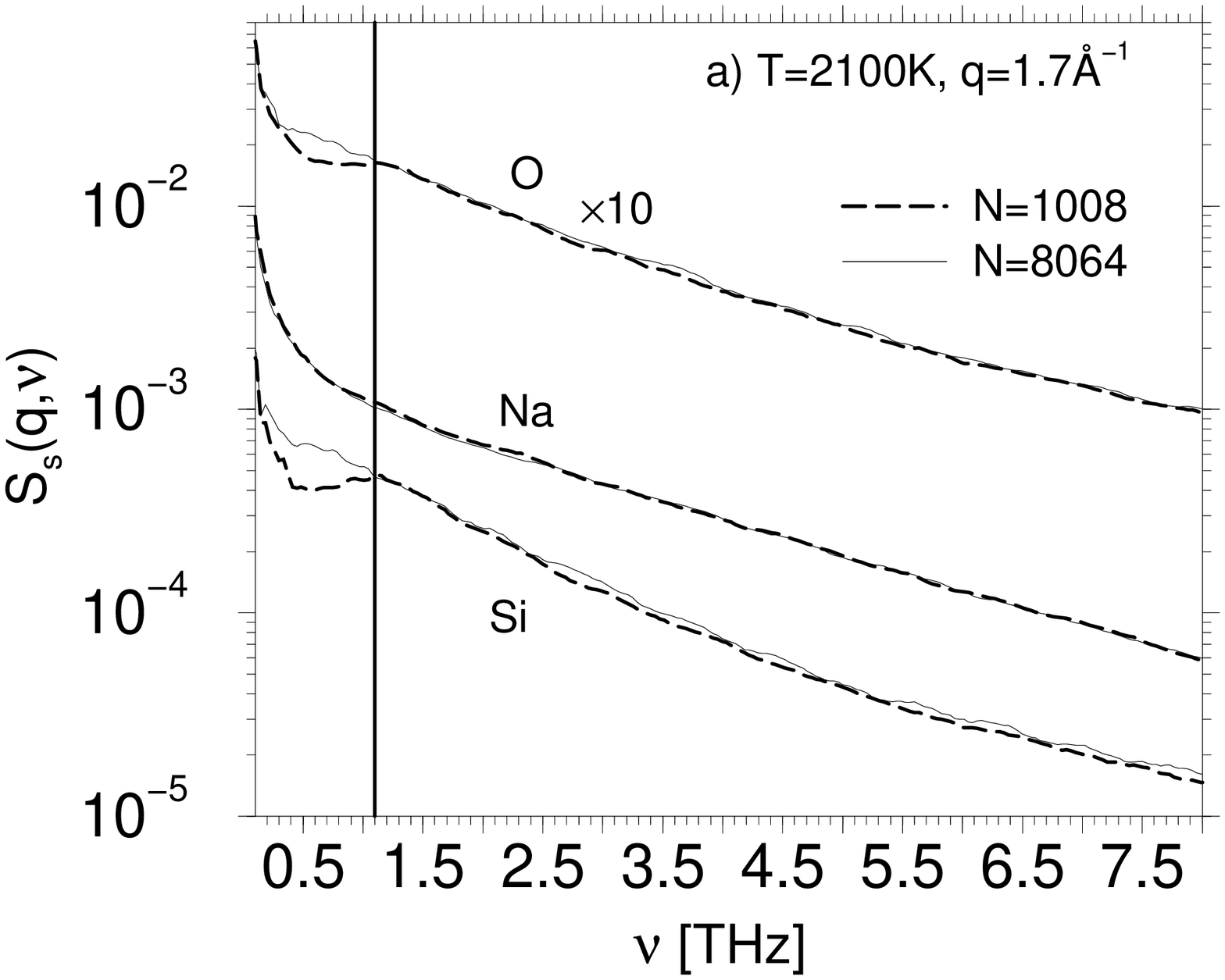,width=13cm,height=9.5cm}
\psfig{file=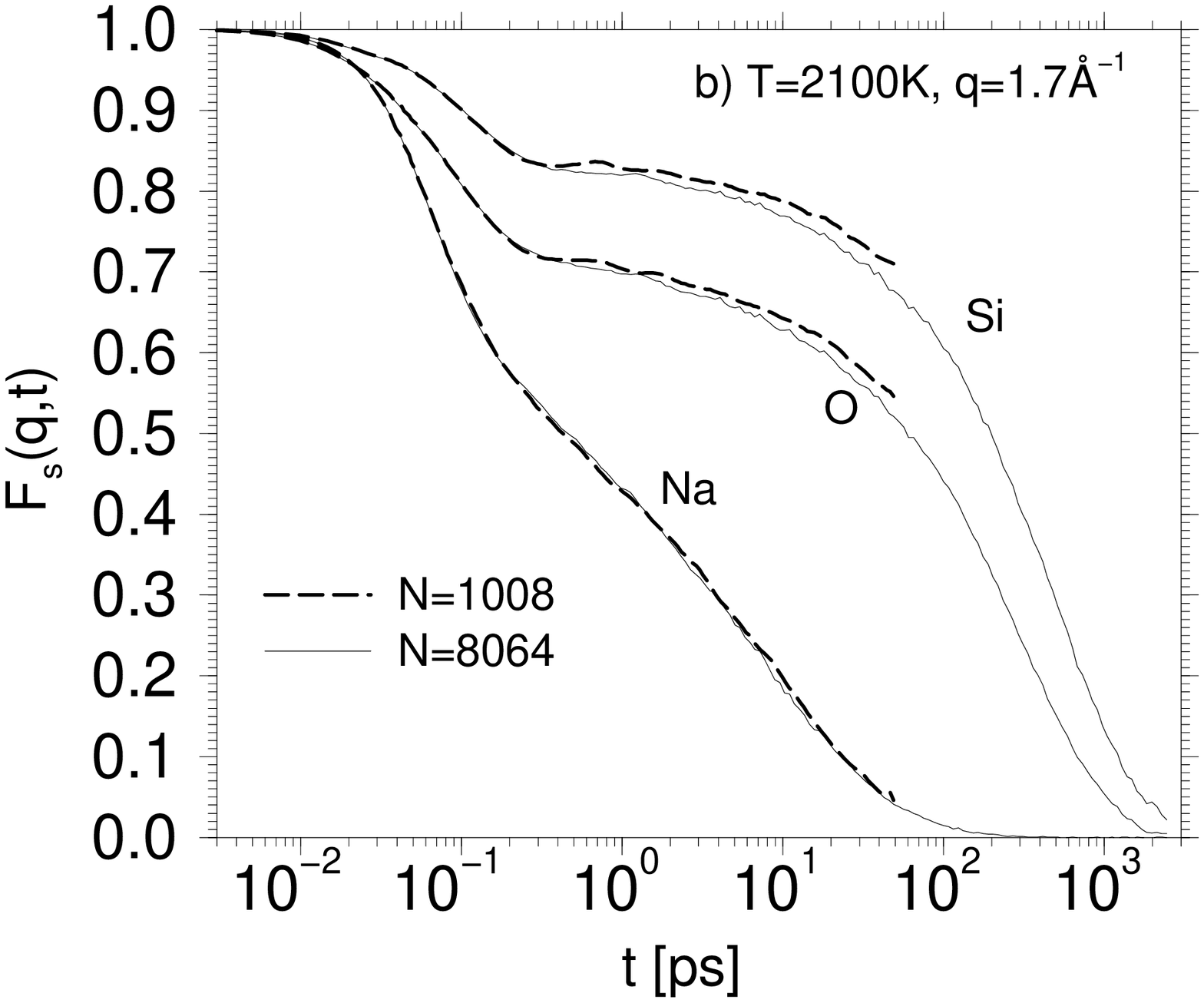,width=13cm,height=9.5cm}
\caption{a) Self part of the dynamic structure factor for silicon,
	 sodium and oxygen for the system sizes $N=1008$ and 
	 $N=8064$ at the temperature $T=2100$~K and the 
	 wave--vector $q=1.7$ \AA$^{-1}$. The vertical line at 
	 $\nu=1.1$ THz marks the frequency below which there is a loss
	 of intensity in the small system. b) Incoherent intermediate
	 scattering function $F_{{\rm s}}(q,t)$ under the same
	 conditions as in a).\label{fig3}}
\end{figure}

Having described the structure of SDS we turn now our attention to
a dynamical quantity, namely the self part of the dynamic structure
factor $S_{{\rm s}} (q, \nu)$, which depends on frequency $\nu$ and the
magnitude of the wave--vector $q$. It is defined by
\begin{eqnarray}
 S_{{\rm s}}(q,\nu) & = & \frac{N_{\alpha}}{N} \int_{-\infty}^{\infty}
                       dt \; {\rm e}^{-2 \pi \nu t} 
                         \left< 
                            {\rm e}^{i {\bf q} \cdot 
                               [{\bf r}_{\alpha}(t) - 
                       {\bf r}_{\alpha}(0)]} \right> \nonumber \\
                    & = & \frac{N_{\alpha}}{N} \int_{-\infty}^{\infty}
                      dt \; {\rm e}^{-2 \pi \nu t} F_{{\rm s}}(q,t)
                    \quad \quad \quad \quad 
		    \alpha \in [{\rm Si, Na, O}]
 \label{eq4} 
\end{eqnarray}
where ${\bf r}_{\alpha}(t)$ is the position vector of a tagged particle
of type $\alpha$ at time $t$ and $F_{{\rm s}}(q,t)$ is the incoherent
intermediate scattering function.  The details of the Fourier
transformation in (\ref{eq4}) are given elsewhere (Horbach {\it et
al.}~1999a).  $S_{{\rm s}}(q,\nu)$ for silicon, sodium and oxygen is
shown in Fig.~\ref{fig3}a at the temperature $T=2100$ K and the
wave--vector $q=1.7$ \AA$^{-1}$ for the two system sizes $N=1008$ and
$N=8064$.  For sodium the curves for the small and the large system
coincide over the whole frequency range. This is not the case for
silicon and oxygen for which $S_{{\rm s}}(q, \nu)$ 
has no system size dependence for frequencies $\nu \ge 1.1$~THz
whereas for smaller frequencies there is a missing of intensity for the
small system. The vibrational modes causing a shoulder around $\nu =
0.9$ THz in $S_{{\rm s}}(q, \nu)$ are usually called boson peak
excitations. In the small system a part of these excitations is missing
due to the loss of intensity for $\nu \le 1.1$ THz.  An explanation of
this behavior for the case of silica, which shows qualitatively the
same finite size effects, can be found in Horbach {\it et al.} (1999a
and 1999b).  Due to the sum rule $\int d \nu \; S_{{\rm s}}(q,\nu) = 1$
the missing of the boson peak excitations for frequencies 0.4 THz $\le
\nu \le$ 1.1 THz in the small system has to be ``reshuffled'' to
smaller frequencies leading to a broadening and an increase of the
quasielastic line around $\nu = 0$. Since the quasielastic line is
outside the frequency resolution of our Fourier transformation the
consequences in the change of the quasielastic line can be observed
better in the Fourier transform of $S_{{\rm s}}(q,\nu)$, i.e.~the
incoherent intermediate scattering function $F_{{\rm s}}(q,t)$, which
is shown in Fig.~\ref{fig3}b for the system sizes $N=8064$ and $N=1008$
at $q=1.7$ \AA$^{-1}$. We recognize from this figure that $F_{{\rm
s}}(q,t)$ shows a two step relaxation behavior similar to the case of
silica and fragile glassformers (Ngai, Riande, and Ingram 1998). As
expected from our results for $S_{{\rm s}}(q,\nu)$ the scattering
functions $F_{{\rm s}}(q,t)$ have no system size dependence for
sodium.  In $F_{{\rm s}}(q,t)$ for silicon and oxygen the height of the
plateau increases and the $\alpha$ relaxation process shifts to longer
times with decreasing system size.  Furthermore, the scattering
functions for the small system show a pronounced oscillation for
$t>0.2$ ps which is due to the fact that the boson peak excitations
present in the small system cause a peak in $S_{{\rm s}}(q,\nu)$
whereas in the large system only a shoulder is present. Finally we
mention that the finite size effects in the dynamics of SDS are found
in the whole $q$ range and, moreover, do not affect the static
properties of SDS.

{\bf Acknowledgments:}

This work was supported by SFB 262/D1 and by Deutsch--Israelisches 
Projekt No.~352--101. We also thank the RUS in Stuttgart for a
generous grant of computer time on the Cray T3E.

\newpage
\section*{References}
\begin{trivlist}
\item[]
Bacon, G. E., 1972, {\it Acta Cryst. A}, {\bf 28}, 357.
\item[]
Cormack, A. N., and Cao, Y., in {\it Modelling of Minerals and
Silicated Materials}, Eds.: B. Silvi and P. Arco (Kluwer,
Dordrecht 1997).
\item[]
Horbach, J., Kob, W., Binder, K., and Angell C. A., 1996,
{\it Phys. Rev. E}, {\bf 54}, R5889.
\item[]
Horbach, J., Kob, W., and Binder, K., 1999a (submitted to Phys. Rev. B)
\item[]
Horbach, J., Kob, W., and Binder, K., 1999b, to appear in the 
Proceedings on {\it Neutrons and Numerical Methods, Grenoble, 1998}, 
preprint in cond-mat/9901162
\item[]
Horbach, J., Kob, W., and Binder, K., 1999c (to be published)
\item[]
Kramer, G. J., de Man, A. J. M., and van Santen, R. A., 1991,
{\it J. Am. Chem. Soc.}, {\bf 113}, 6435.
\item[]
Misawa, M., Price, D. L., and Suzuki, K., 1980,
{\it J. Non--Cryst. Solids}, {\bf 37}, 85.
\item[]
Ngai, K. L., Riande, E., and Ingram, M.D., (editors), 1998,
{\it J. Non--Cryst. Solids}, {\bf 235--237} (Proceedings of the
{\it Third International Discussion Meeting on Relaxations in 
Complex Systems}, Vigo, 1997).
\item[]
Smith, W., Greaves, G. N., and Gillan, M. J., 1995, 
{\it J. Chem. Phys.}, {\bf 103}, 3091.
\item[]
Susman, S., Volin, K. J., Montague, D. G., and Price, D. L., 1991,
{\it Phys. Rev. B}, {\bf 43}, 11076.
\item[]
van Beest B. W., Kramer G. J., and van Santen R. A., 1990,
{\it Phys. Rev. Lett.}, {\bf 64} 1955.
\item[]
Vessal, B., Amini, A., Fincham, D., and Catlow, C. R. A., 1989,
{\it Philos. Mag. B}, {\bf 60}, 753.
\end{trivlist}

\newpage

\end{document}